\documentclass[aps,twocolumn, secnumarabic,nobalancelastpage,amsmath,amssymb,nofootinbib]{revtex4-2}

\usepackage[utf8]{inputenc}
\usepackage[caption=false]{subfig}
\usepackage{ragged2e}
\DeclareCaptionJustification{justified}{\justifying}
\usepackage{graphicx}
\usepackage{bm}
\usepackage{bbm}
\usepackage{bbold}
\usepackage{amsmath}
\usepackage{braket}
\usepackage[normalem]{ulem}
\usepackage{soul}
\usepackage{txfonts}
\usepackage{tabularx}
\usepackage{tabularray}
\usepackage[usenames,dvipsnames]{color}
\setstcolor{red}

\usepackage{geometry}
\usepackage[colorlinks=true, linkcolor=blue, citecolor=blue, urlcolor=blue]{hyperref}

\newlength\myboxwidth
\begin{document}

\title{Material Loss Model Calibration for Tantalum Superconducting Resonators}

\author{Guy Moshel$^{1}$}
\email{gmoshel@campus.technion.ac.il}
\author{Sergei Masis$^{2}$}
\author{Moshe Schechter$^{3}$}
\author{Shay Hacohen-Gourgy$^{1}$}

\affiliation{$^{1}$ Department of Physics, Technion - Israel Institute of Technology, Haifa 32000, Israel}
\affiliation{$^{2}$ Physikalisches Institut, Karlsruher Institut für Technologie}
\affiliation{$^{3}$ Department of Physics, Ben-Gurion University of the Negev, Beer Sheva 84105, Israel}

\date{\today }
\begin{abstract}
Material research is a key frontier in advancing superconducting qubit and circuit performance. In this work, we develop a simple and broadly applicable framework for accurately characterizing two-level system (TLS) loss using internal quality factor measurements of superconducting transmission line resonators over a range of temperatures and readout powers. We applied this method to a series of $\alpha$-Ta resonators that span a wide frequency range, thus providing a methodology for probing the loss mechanisms in the fabrication process of this emerging material for superconducting quantum circuits. We introduce an analytical model that captures the loss behavior without relying on numerical simulations, enabling straightforward interpretation and calibration. Additionally, our measurements reveal empirical frequency-dependent trends in key parameters of the model, suggesting contributions from mechanisms beyond the standard tunneling model of TLSs. 
\end{abstract}
\maketitle

\section{Introduction}

Superconducting qubits and resonators are the cornerstone of many advanced quantum information systems, including quantum simulations, quantum error correction, and quantum communication~\cite{devoret2013superconducting,wendin2017quantum,houck2012chip}. Despite significant progress, scaling quantum processors to practical sizes remains constrained by qubit coherence times, which is currently limited by losses that far exceed those anticipated from the bulk properties of the constituent materials~\cite{siddiqi2021engineering}. These elevated losses suggest that critical decoherence pathways arise from uncontrolled surfaces, interfaces, and contaminants. Substantial work has been done to reduce these interfacial losses, in various material platforms~\cite{biznarova2024mitigation,melville2020comparison,deng2023titanium}. Specifically, tantalum-based superconducting qubits have achieved record lifetimes, with coherence times exceeding 0.3ms~\cite{place2021new,wang2022towards}. These results have been reproduced across diverse fabrication methods and substrates, underscoring the potential of tantalum as an important material for quantum hardware. Nevertheless, the sources of loss in high-quality-factor (Q-factor) tantalum devices remain poorly understood and present a major obstacle to further improvement. A detailed and accurate model for the loss mechanisms in tantalum is necessary.

The primary loss mechanisms in superconducting devices are radiation and packaging ~\cite{huang2021microwave,sage2011study}, non-equilibrium quasiparticles~\cite{serniak2018hot}, bulk material absorption~\cite{braginsky1987experimental,gurevich1991intrinsic} and two-level systems (TLSs) on the material surfaces~\cite{mcrae2020materials}. The contribution of TLSs is difficult to differentiate and accurately estimate because of the presence of multiple TLS sources within a single device and their dependence on the device geometry and fabrication method. Work had been done on various materials~\cite{burnett2018noise,burnett2015analysis,paik2010reducing,lisenfeld2019electric}, but tantalum remained under-researched. 

In this work, we systematically investigate the contributions of TLSs and other loss mechanisms in tantalum superconducting resonators. By varying the readout power by many orders of magnitude, and also the temperature, we achieve precise characterization of subtle differences between the loss mechanisms in the resonators. We expand on the work of~\cite{crowley2023disentangling}, which used a similar saturation model but with added empirical parameters. Instead, we incorporate the spatial variation of the electric field in the resonators into our model, which improved the fit quality without adding any parameters to the model.

In addition, we measured resonators across a wide frequency range: from 2.3GHz to 15.7GHz. This allowed us to observe frequency trends in the model parameters that were extracted for each resonator, shedding light on the underlying physics of the TLS loss, specifically the frequency dependence of the density of states.

\section{setup}

\begin{figure}[htp!]
    \centering
    \includegraphics[width=\linewidth, trim=0 0 0 0, clip]{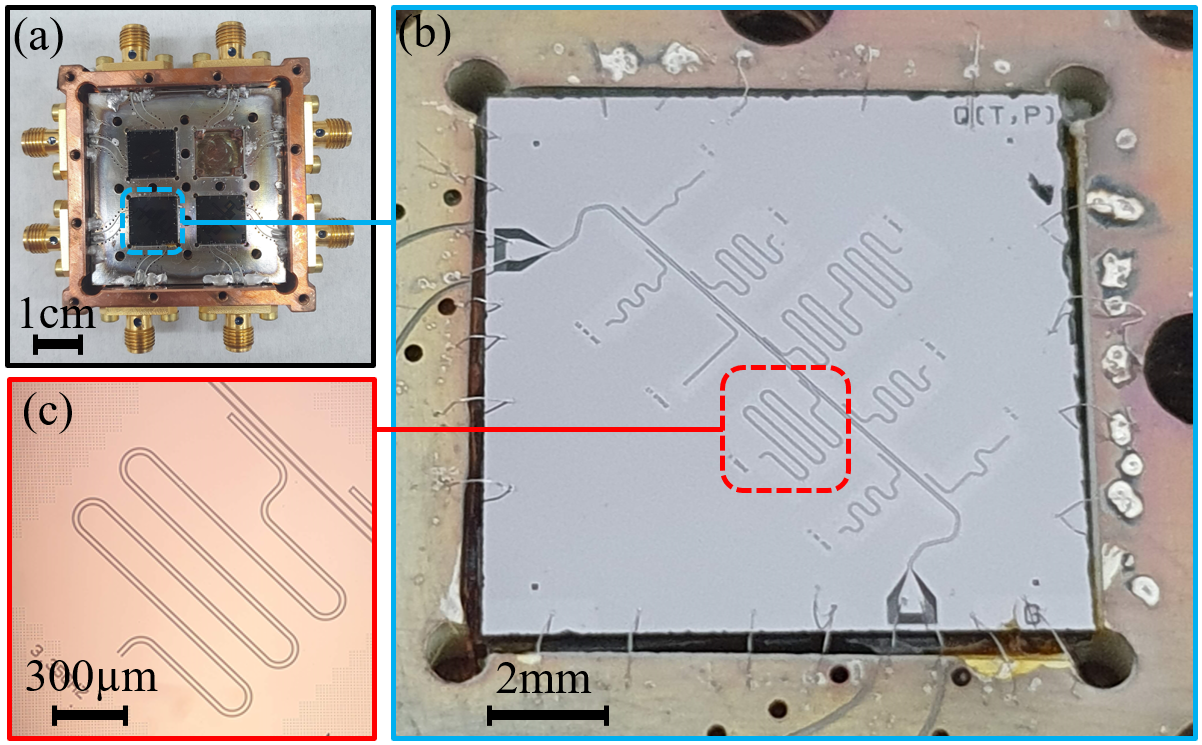}
    \caption{The experimental setup. (a) The copper box and silver-coated PCB for chip mounting. Only the marked slot was used in this work. (b) The fabricated chip after mounting. Bright areas are tantalum and dark areas are sapphire substrate (etched tantalum). wire bond between the tantalum and PCB are visible. (c) A micrograph of one of the measured resonators.}
    \label{fig:chip}
\end{figure}

\begin{figure}[htp!]
    \centering
    \includegraphics[width=\linewidth, trim=0 0 0 0, clip]{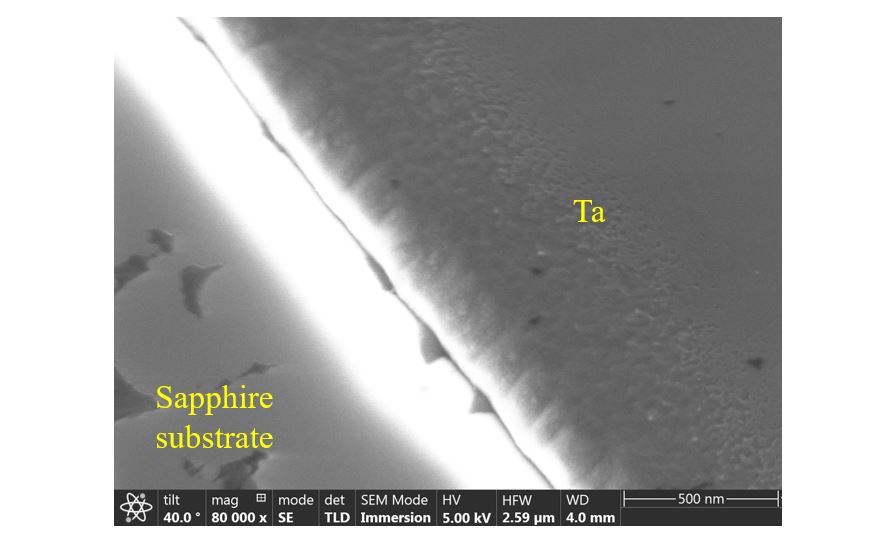}
    \caption{A SEM image of an edge of the etched tantalum, in the optimized wet process. The smooth boundary indicates valid etching. The white strip is an imaging artifact caused by substrate charging.}
    \label{fig:SEM}
\end{figure}

We used a 2-inch, 420$\mu$m thick sapphire wafer, sputtered with a 170nm thick $\alpha$-Ta layer by STAR cryoelectronics. The wafer was cleaned by sequential 5 minute sonication in acetone, methanol, and isopropanol, and then coated with AZ1518 photo-resist, exposed with the desired pattern using a DWL66+ laser writer by Heidelberg, and developed in TMAH (tetramethylammonium hydroxid) for 40 seconds. A hard bake was then done, and the tantalum was wet-etched in Ta-etchant 111 by Transene for 21 seconds. We examined the smoothness of the edges of the resulting tantalum structures using a scanning electron microscope (SEM), according to the characteristics in~\cite{place2021new}. The etching results are shown in Fig.~\ref{fig:SEM}. The wafer was then diced into the correct chip size and mounted on a designated box made of oxygen-free high-thermal-conductivity (OFHC) copper. The chip was placed on four copper pedestals at its corners and secured using ge varnish, which is thermally conductive at cryogenic temperatures. A silver-coated printed circuit board (PCB) was electrically and thermally anchored to the copper box using indium solder. The ground and signal ports of the chip were electrically connected to the PCB using aluminum wire bonds, as shown in Fig.~\ref{fig:chip}. Finally, the box was sealed with an OFHC copper lid and mounted inside the 10mK stage of a commercial BlueFors dilution cryostat. Care was taken to ensure good thermal anchoring to the cryostat. A $\mu$-metal shield was used to repel stray magnetic fields from the sample at low temperatures. 

The patterned structures were a series of $\lambda/4$ transmission line resonators, coupled to a common feedline in a hanging (or notch) configuration, as shown in Fig.~\ref{fig:chip}(c). The resonators and feedline were made of coplanar waveguide (CPW) segments. \\

We measured the complex transmission signal $S_{12}$ through the feedline to find the full lineshapes of the base mode of each resonator. By fitting each lineshape according to the procedure laid out in~\cite{probst2015efficient} using the qkit python module, we extracted the resonance frequency and the internal and external Q-factors ($Q_i$ and $Q_e$, respectively). The measurements were carried out using a Keysight network analyzer, PNA-L model N5232B. For the temperature scans, we waited 30 minutes at each temperature point for the system to thermalize, and only then performed the power scans. We calculated the microwave power inside the feedline from the known input power using the nominal attenuator values at cryogenic temperature. The average microwave energy stored in the resonators for an impedance matched feedline, is given by~\cite{mcrae2020materials} 
\begin{equation}\label{eq:n_avg}
    W = \frac{2 Q_l^2}{Q_e} \frac{P_0}{\omega_0},
\end{equation}
where $Q_l$ and $Q_e$ are the loaded and external quality factors of the resonator, respectively,  $\omega_0$ is its angular resonant frequency and $P_0$ is the power in the feedline. To convert this energy to voltage, which we used in the model described below, we used
\begin{equation}\label{eq:n_avg}
    V_0 = \sqrt{2 W Z_0},
\end{equation}
where $Z_0=50\Omega $ is the resonator impedance.

\section{model}

We begin by calculating the electric field distribution in the transmission line resonators. In the base mode of a $\lambda /4$ CPW transmission line resonator, the voltage and current profiles are~\cite{pozar2021microwave} 
\begin{equation}\label{eq:Vz}
V(z)=V_0\sin{\left(\frac{2\pi}{\lambda}z\right)},\;\;\; I(z)=\frac{V(z)}{Z_0},
\end{equation}
were $Z_0=\sqrt{L/C}$ is the impedance of the transmission line and $C$ and $L$ are the capacitance and inductance per unit length, respectively. $V_0$ is the voltage amplitude of the mode excitation. The average stored energy in the resonator per cycle is
\begin{equation}
W =\int_0^l{dz\left(\frac{C V^2(z)}{2}+\frac{L I^2(z)}{2}\right)} = \frac{\lambda C V_0^2}{8}.
\end{equation}
The loaded quality factor of the resonator is given by
\begin{equation}
    Q_{l}=\omega_0\frac{W}{P},
\end{equation}
where $P$ is the energy loss per cycle and $\omega_0=1/\sqrt{LC}$ is the mode resonant angular frequency. $Q_l$ can be separated into different loss channels,
\begin{equation}\label{eq:Ql_Qe_Qi}
 \frac{1}{Q_l}=\frac{1}{Q_e}+\frac{1}{Q_i}=\frac{1}{Q_e}+\frac{1}{\omega_0}\sum_j{\frac{P_j}{W}}.
\end{equation}
The external quality factor, $Q_e$, represents the energy loss to the feedline, which can be measured by the experimenter.  The internal quality factor, $Q_i$, represents all the other loss channels, which cannot be measured individually. $P_j$ is the power loss in the $j$-th channel, where $j$ goes over all possible internal loss channels. The three main loss channels for thin-film superconducting transmission line resonators are two-level systems (TLSs) excitation loss, quasiparticles (QPs) dissipation in the superconductor, and other mechanisms which are constant in temperature and power, such as radiation loss~\cite{mcrae2020materials,grunhaupt2018loss,dial2016bulk,murray2021material,gambetta2016investigating}. We can write
\begin{equation}\label{eq:Qi_terms}
\frac{1}{Q_i} = \frac{P_{\mathrm{TLS}} + P_{\mathrm{qp}} + P_{\mathrm{other}}}{\omega_0 W} 
= \frac{1}{Q_{\mathrm{TLS}}} + \frac{1}{Q_{\mathrm{qp}}} + \frac{1}{Q_{\mathrm{other}}}.
\end{equation}

$Q_{\mathrm{other}}$ mostly depends on the electromagnetic environment of the resonator and not on the temperature or stored energy (equivalent to drive power)~\cite{crowley2023disentangling}.
$Q_{\mathrm{qp}}$ is also independent of power but does depend on temperature. the QP loss can be derived from the complex surface impedance of thin-layer tantalum, $Z_{\mathrm{s}} = R_{\mathrm{s}} + iX_{\mathrm{s}}$, through the relation
\begin{equation}\label{eq:Qqp}
    Q_{\mathrm{qp}} = \frac{X_{\mathrm{s}}}{R_{\mathrm{s}}}.
\end{equation}
At temperatures sufficiently below $T_\mathrm{c}$ we can approximate~\cite{mcrae2020materials}
\begin{equation}\label{eq:Rs}
    R_{\mathrm{s}} = \frac{1}{2}\omega_0^2 \mu_0^2 \lambda^3 \sigma_1,
\end{equation}
and
\begin{equation}\label{eq:Xs}
    X_{\mathrm{s}} = \omega_0\mu_0\lambda,
\end{equation}
where $\lambda$ is the London penetration depth and $\sigma_1$ is the real part of the complex conductivity of tantalum. An expression for $\sigma_1$, valid in the limit $k_{\mathrm{B}}T \ll \hbar\omega \ll \Delta$, with $\Delta$ being the superconducting gap energy, is~\cite{mazin2005microwave}
\begin{equation}\label{eq:sigma_1}
    \frac{\sigma_1}{\sigma_{\mathrm{n}}} = \frac{2\Delta}{\hbar\omega}\exp{\left(-\frac{\Delta}{k_{\mathrm{B}}T}\right)}K_0\left(\frac{\hbar\omega}{2k_{\mathrm{B}}T} \right)\sinh\left(\frac{\hbar\omega}{2k_{\mathrm{B}}T} \right),
\end{equation}
where $K_0(x)$ is the modified Bessel function of the second kind of order zero, and $\sigma_{\mathrm{n}}$ is the normal state conductivity just above the critical temperature. Substituting Eqs.~\ref{eq:Rs},~\ref{eq:Xs} and ~\ref{eq:sigma_1} into Eq.~\ref{eq:Qqp} yields
\begin{equation}\label{eq:Qqp_full}
    Q_{\mathrm{qp}} = Q_0^{\mathrm{qp}}\frac{\exp{\left(\frac{\Delta}{k_{\mathrm{B}}T}\right)}}{K_0\left(\frac{\hbar\omega}{2k_{\mathrm{B}}T} \right)\sinh\left(\frac{\hbar\omega}{2k_{\mathrm{B}}T} \right)},
\end{equation}
Were we defined $Q_0^{\mathrm{qp}}\equiv \hbar/\sigma_{\mathrm{n}}\mu_0\lambda^2\Delta$ which we treat as a fitting parameter in subsequent sections. This is similar to the approach of~\cite{crowley2023disentangling}. Notice that Eq.~\ref{eq:Qqp_full} explicitly contains the dependence on the resonator frequency, as well as the temperature dependence. This is essential for explaining results from a wide range of resonator frequencies, as in the present work. The slight temperature dependence of $\Delta$ and $\lambda$ is negligible. It is worth noting that Eq.~\ref{eq:Qqp_full} assumes thermal equilibrium of QPs in the resonators, which is not necessarily the case, as external disturbances can significantly change the energy distribution in the superconductor~\cite{de2011number,serniak2018hot,connolly2024coexistence}. Common disturbances include infrared and terahertz photons that leak into the cryopackage and are absorbed by the chip, background radioactive radiation from the environment, and impacts of cosmic particles. It was also proposed that interaction between TLSs and QPs can change the steady-state energy distribution~\cite{de2020two}. In general, these effects result in a constant reduction in $Q_i$, which can be absorbed into $Q_{\mathrm{other}}$.

$Q_{\mathrm{TLS}}$ depends on temperature, power and resonator frequency. Here, we develop a model for $P_{\mathrm{TLS}}$, which we will use to calculate $Q_{\mathrm{TLS}}$. Consider an electric field distribution $\vec{E}(\vec{r})$ in a lossy dielectric material of volume $\mathcal{V}$, loss tangent $\tan{\delta}$ and dielectric constant $\varepsilon$. The dissipated power in the dielectric is
\begin{equation}\label{eq:P_definition}
P=\varepsilon \int_{\mathcal{V}}  \tan{\delta} \left|\vec{E}(\vec{r})\right|^2 d\mathcal{V}.
\end{equation}
The loss tangent for TLSs is given by the standard model for TLSs~\cite{von1977saturation,martinis2005decoherence,pappas2011two}
\begin{equation}\label{eq:tan_deltaj}
\tan{\delta}=\delta^0\frac{\tanh{\left(\frac{U}{2k_{\mathrm{B}}T}\right)}}{\sqrt{1+|\vec{E}(\vec{r})|^2/E_{\mathrm{s}}^2}},
\end{equation}
where $U$ is the TLS transition energy and $E_\mathrm{s}$ is the TLS saturation field. The integration over the volume of the dielectric should take into account the dependence of the loss tangent on $\vec{E}(\vec{r})$. The saturation field is proportional to $\sqrt{\Gamma_1 \Gamma_2}$~\cite{tai2024anomalous,gao2008physics}, where
\begin{equation} \label{eq:Q_TLS_defs}
\begin{split}
    &\Gamma_2 = \frac{1}{T_2} \propto k_{\mathrm{B}} T, \\
    &\Gamma_1 = \frac{1}{T_1} \propto \frac{U}{\tanh{(U/2k_{\mathrm{B}} T)}}, \\
\end{split}    
\end{equation}
and therefore
\begin{equation}\label{eq:Es}
E_{\mathrm{s}}=\sqrt{\frac{T}{\zeta}\frac{U/k_{\mathrm{B}}}{\tanh{(U/2k_{\mathrm{B}}T)}}},
\end{equation}
where $\zeta$ is a material-dependent constant. We see that the power saturation of the TLSs depends on temperature. Since most of the energy stored in the resonator is at the resonance frequency $\omega$, only TLSs with this transition frequency are relevant for power loss. We therefore assume $U=\hbar\omega$. 

Following~\cite{martinis2022surface}, we assume that most of the dielectric loss occurs on the three interfaces of the resonator: metal-air (MA), substrate-air (SA) and metal-substrate (MS). Each interface is modeled as a dielectric layer with thickness $t_j$, relative dielectric constant $\varepsilon_j$ and loss tangent $\tan \delta_j$, where $j\in\{\mathrm{MA,SA,MS}\}$ is the subscript of the interface. Since the thicknesses of the interfaces are on the order of a single nanometer~\cite{woods2019determining}, a reasonable assumption is that the field does not change direction and magnitude within the interface. We can therefore use the boundary conditions for the electric field at the interface between two dielectrics, assuming that no free charges exist, to approximate the electric fields inside the films~\cite{wenner2011surface}. We obtain
\begin{equation}\label{eq:EWenner}
\begin{split}
&\vec{E}_{\mathrm{MA}} \approx \hat{n}_{\perp} (\vec{E}_{\mathrm{A}}\cdot \hat{n}_{\perp})/\varepsilon_{\mathrm{MA}},\\
&\vec{E}_{\mathrm{MS}} \approx \hat{n}_{\perp} (\vec{E}_{\mathrm{S}} \cdot \hat{n}_{\perp} ) \varepsilon_{\mathrm{S}}/ \varepsilon_{\mathrm{MS}},\\
&\vec{E}_{\mathrm{SA}} \approx \hat{n}_{\perp} (\vec{E}_{\mathrm{A}}\cdot \hat{n}_{\perp}) / \varepsilon_{\mathrm{SA}}+\hat{n}_{\parallel}(\vec{E}_{\mathrm{A}}\cdot \hat{n}_{\parallel}).
\end{split}
\end{equation}
Here, $\vec{E}_{\mathrm{A}}$ and $\vec{E}_{\mathrm{S}}$ are the electric fields in the air and in the substrate immediately outside of the interface layers, respectively. Unit vectors $\hat{n}_{\perp}$ and $\hat{n}_{\parallel}$ denote the perpendicular and parallel directions relative to the interface surface, respectively.

\begin{figure}[htp!]
    \centering
    \includegraphics[width=\linewidth, trim=0 20 0 20, clip]{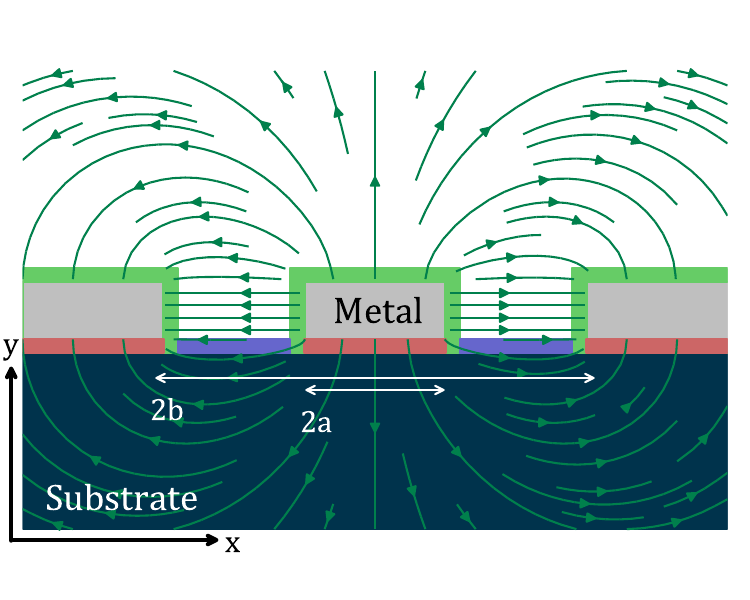}
    \caption{TEM mode electric field of a CPW in the transverse plane, according to Eq.~\ref{eq:EMurray1}. The metal-air (MA) interface is marked in green, the metal-substrate (MS) interface is marked in red and the substrate-air (SA) interface is marked in blue. Note that in Eq.~\ref{eq:EMurray1}, $x$ is measured from the axis of symmetry and $y$ from the metal surface.}
    \label{fig:CPW_Efield}
\end{figure}

The electric field in the TEM mode of the CPW can be analytically calculated under a quasi-static approximation, which requires the wavelength to be much larger than the cross-sectional width of the CPW. Assuming a thin metal layer and a semi-infinite substrate, under a voltage $V$ between the centerline and the ground planes, the field components are given by~\cite{murray2018analytical}:
\begin{equation}\label{eq:EMurray1}
\begin{split}
iE_{\mathrm{x}}(w,z)+&E_{\mathrm{y}}(w,z) =  \\
 &V(z)\frac{ g_{\mathrm{a,b}}(w)}{ K(k')}\sqrt{\frac{b^2}{(w^2-a^2)(w^2-b^2)}},
\end{split}
\end{equation}
where $2a$ is the width of the central conductor, $2b$ is the distance between the ground planes, $w=x+iy$ is the complex planar coordinate, $K(x)$ is the complete elliptic integral of the first kind and $k'=\sqrt{1-(a/b)^2}$. $V(z)$ is given by Eq.~\ref{eq:Vz}. The sign function $g_{\mathrm{a,b}}(w)$ is defined by
\begin{equation}\label{eq:EMurray2}
g_{\mathrm{a,b}}(w)= 
\begin{cases}
1 & \text{if \;\;} y^2-x^2 \leq \frac{a^2+b^2}{2}\\
-1 & \text{otherwise }
\end{cases}
\end{equation}
The field is depicted in Fig.~\ref{fig:CPW_Efield}, superimposed on a qualitative cross-sectional diagram of the CPW. Further analytical corrections are available for metal corners~\cite{martinis2022surface} and substrate trenches~\cite{murray2020analytical}, although we did not use these in this work. 

The fields in the lossy interfaces are found by substituting the fields immediately outside of the interfaces from Eq.~\ref{eq:EMurray1} into Eq.~\ref{eq:EWenner}. For this we take $\hat{n}_{\perp}=\hat{y}$ and $\hat{n}_{\parallel}=\hat{x}$. Note that if we use $y=0$ at the interfaces, the field diverges at $x=a,b$. We therefore used $y=t_j$. We can now substitute for each interface the resulting fields in Eq.~\ref{eq:P_definition} and obtain the power loss $P_j$ for that interface. Summing over all interfaces gives us $P_\mathrm{TLS}$, from which we can calculate $Q_\mathrm{TLS}$ using Eq.~\ref{eq:Qi_terms}. Overall we obtain a function $Q_\mathrm{TLS}(T,V_0)$ that gives $Q_{\mathrm{TLS}}$ as a function of temperature and voltage amplitude, given the material parameters $t_j$, $\zeta_j$, $\delta_j^0$ and $\varepsilon_j$ for all interfaces, and the CPW dimensions $a$ and $b$.

Certain assumptions can be made to greatly reduce the number of free parameters in the model. First, the electric field magnitude $|\vec{E}(\vec{r})|^2$ in the SA interface is negligible compared to that of the MS and MA interfaces, so its loss can be neglected in the calculation of $P_{\mathrm{TLS}}$. Second, since the field profiles inside the MS and MA interfaces are identical, we have no way of distinguishing between their contributions to $P_{\mathrm{TLS}}$. We assume that the loss in the MA interface is negligible compared to the loss in the MS interface. This leaves us with only four model parameters: $t_{\mathrm{MS}}$, $\zeta_{\mathrm{MS}}$, $\delta_{\mathrm{MS}}^0$ and $\varepsilon_{\mathrm{MS}}$. The MS layer is made of amorphous $\mathrm{AlO_x}$ or amorphous tantalum. Based on typical native oxide thicknesses of tantalum, we used $t_{\mathrm{MS}} = 2nm$~\cite{mclellan2023chemical}. For the dielectric constant, we took $\varepsilon_{\mathrm{MS}} = 15$~\cite{sangwan2024exploration}. This leaves only two parameters in the model. 

We now write the expression for $P_{\mathrm{MS}}$ explicitly. Using the axes definitions in Fig.~\ref{fig:CPW_Efield}, the volume integration in Eq.~\ref{eq:P_definition} can be separated into x and y integrations in the cross-sectional plane and z integration along the length of the resonator. Because of the small thickness of the interfacial layer, we can assume that the field does not change appreciably inside of it. This makes the y integration trivial, leaving only the x and z integrations, namely
\begin{widetext}
\begin{equation}\label{eq:MS_integral}
P_{\mathrm{MS}} = \left[2t_{\mathrm{MS}} \frac{\varepsilon_{\mathrm{S}}^2 \delta_{\mathrm{MS}}^0}{\varepsilon_{\mathrm{MS}}} \right] \int_0^{\lambda /4} dz \left( \int_0^a dx + \int_b^{\infty} dx \right) 
 \frac{\left[ \frac{V_0^2 b^2}{K^2(k')} \tanh{\frac{\hbar \omega}{2 k_{\mathrm{B}} T}} \right] \sin^2{\left(\frac{2\pi}{\lambda} z\right)} f_y^2(x) }{\sqrt{ 1 + \frac{1}{T}\frac{k_{\mathrm{B}}\zeta_{\mathrm{MS}} \varepsilon_{\mathrm{S}}^2 }{\hbar\omega \varepsilon_{\mathrm{MS}}^2} \left[ \frac{V_0^2 b^2}{K^2(k')} \tanh{\frac{\hbar \omega}{2 k_{\mathrm{B}} T}} \right] \sin^2{\left(\frac{2\pi}{\lambda}z\right)} f_y^2(x) }},
\end{equation}
\end{widetext}
where we defined the function 
\begin{equation}\label{eq:fx}
f_y(x)  \equiv  \mathrm{Re} \left\{ \frac{1}{\sqrt{(x^2-a^2)(x^2-b^2)}} \right\},
\end{equation}
based on Eq.~\ref{eq:EMurray1}. The z integration in Eq.~\ref{eq:MS_integral} can be carried out analytically using the identity
\begin{equation}\label{eq:z_integral_wolframAlpha}
\begin{split}
&\int_0^{\lambda/4} \frac{\sin^2{\left(\frac{2 \pi}{\lambda}z\right)}}{\sqrt{1 + A \sin^2{\left(\frac{2 \pi}{\lambda}z\right)}}} dz \\
& \;\;\;\;\;\;\;\;\;\;\;\;\;\;\;\;\;\;\;\;\;\;\;\;\;\;\;\;\;\;\;\;= \frac{E(\frac{\pi}{2}|-A)-F(\frac{\pi}{2}|-A)}{\frac{2 \pi}{\lambda}A},
\end{split}
\end{equation}
were $A$ is not a function of $z$, the incomplete elliptic integral of the first kind is defined as
\begin{equation}\label{eq:ellipF}
    F(x|m) \equiv \int_0^x \frac{dt}{\sqrt{1-m \sin^2{t}}},
\end{equation}
and the incomplete elliptic integral of the second kind is defined as
\begin{equation}\label{eq:ellipE}
    E(x|m) \equiv \int_0^x \sqrt{1-m \sin^2{t}} dt.
\end{equation}
We are left with only the x integration, which can be performed numerically. The term 
\begin{equation}
    C_{\mathrm{T,V_0}}\equiv \frac{V_0^2 b^2}{K^2(k')} \tanh{\frac{\hbar \omega}{2 k_{\mathrm{B}} T}},
\end{equation}
appearing in Eq.~\ref{eq:MS_integral}, does not depend on material parameters, only on temperature and power. After the analytic $z$ integration Eq.~\ref{eq:MS_integral} becomes

\begin{widetext}
\begin{equation}\label{eq:MS_integral_x}
P_{\mathrm{MS}} = A_{\mathrm{MS}}^0  T  \left( \int_0^a dx + \int_b^{\infty} dx \right)
 \left[ E \left( \frac{\pi}{2} \left| -\frac{1}{T} B_{\mathrm{MS}}^0 C_{\mathrm{T,V_0}} f_y^2(x) \right. \right) \right.
 \left.  - F \left( \frac{\pi}{2} \left| -\frac{1}{T}B_{\mathrm{MS}}^0 C_{\mathrm{T,V_0}} f_y^2(x) \right. \right)
  \right],
\end{equation}
\end{widetext}

where the only material-dependent parameters are 
\begin{equation}\label{eq:A0_ms}
A_{\mathrm{MS}}^0 \equiv \frac{\hbar\omega \lambda}{\pi k_{\mathrm{B}}}\frac{t_{\mathrm{MS}} \varepsilon_{\mathrm{MS}} \delta_{\mathrm{MS}}^0}{\zeta_{\mathrm{MS}}} \;\;\;\; ; \;\;\;\;
B_{\mathrm{MS}}^0 \equiv \frac{k_{\mathrm{B}}}{\hbar\omega}\frac{\zeta_{\mathrm{MS}} \varepsilon_{\mathrm{S}}^2}{\varepsilon_{\mathrm{MS}}^2}.
\end{equation}

Finally, we calculate the total $Q_i$ by substituting Eq.~\ref{eq:Qqp_full} and Eq.~\ref{eq:MS_integral_x} into Eq.~\ref{eq:Qi_terms}. We fit our measurements of $Q_i$ as a function of $T$ and $V_0$ to find the four model parameters $A_{\mathrm{MS}}^0$, $B_{\mathrm{MS}}^0$, $Q_0^{\mathrm{qp}}$ and $Q_{\mathrm{other}}$. This enables us to distinguish the contributions of the different loss channels.

\section{Results and discussion}

\begin{figure}[htp!]
    \centering
    \includegraphics[width=\linewidth, trim=0 10 0 0, clip]{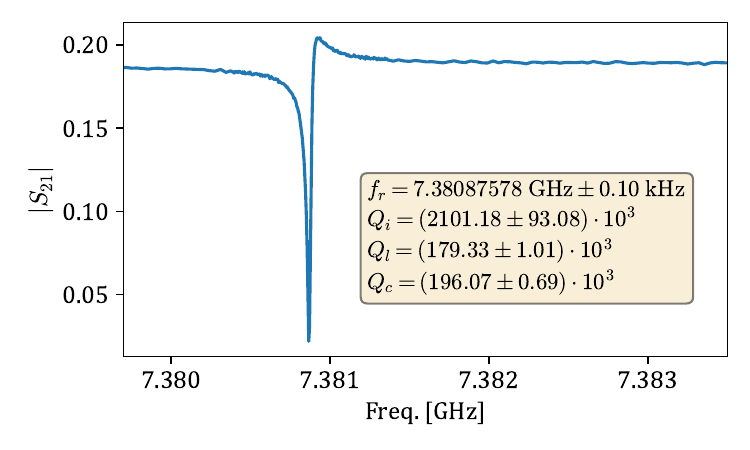}
    \caption{A typical lineshape and extracted Q-factors. This specific lineshape was measured in the 7.38GHz resonator at a temperature of 280mK with a readout power of -30dBm at the entrance of the fridge.}
    \label{fig:lineshape}
\end{figure}

We received high Q-factor for all our resonators ($Q_i \sim 10^6$), with typical measurement features. Some of the resonators shown in Fig.~\ref{fig:chip} were not visible in our measurements, possibly due to the formation of a crack at low temperatures, which created a disconnect somewhere along the resonators. A typical lineshape and the extracted Q-factors are shown in Fig.~\ref{fig:lineshape}. To obtain a high accuracy in $Q_i$, the resonators should have a roughly critical coupling to the feedline. However, this was not always the case, as can be seen in Fig.~\ref{fig:lineshape} for example. The reasons are the difficulty of designing specific $Q_e$ for high-Q resonators, and the variation in $Q_i$ with temperature and readout power. Nonetheless, we were able to get sufficiently accurate measurements of $Q_i$ in all visible resonators. 

\begin{figure*}
    \subfloat{
        \includegraphics[width=.48\linewidth, trim=0 15 0 0, clip]{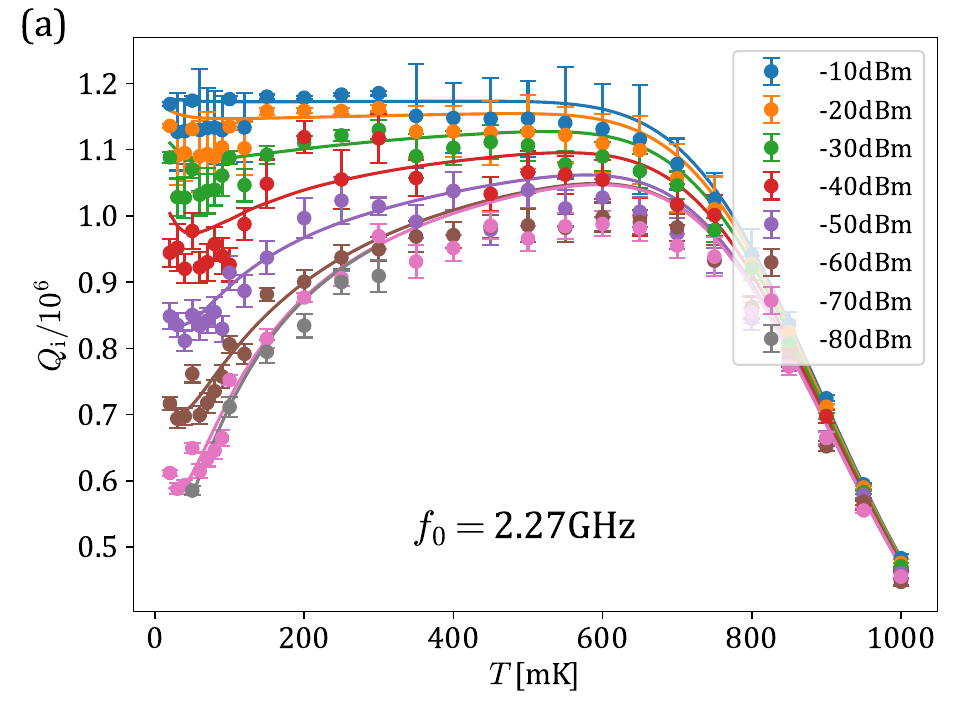}
        }\hfill
    \subfloat{
        \includegraphics[width=.48\linewidth, trim=0 15 0 0, clip]{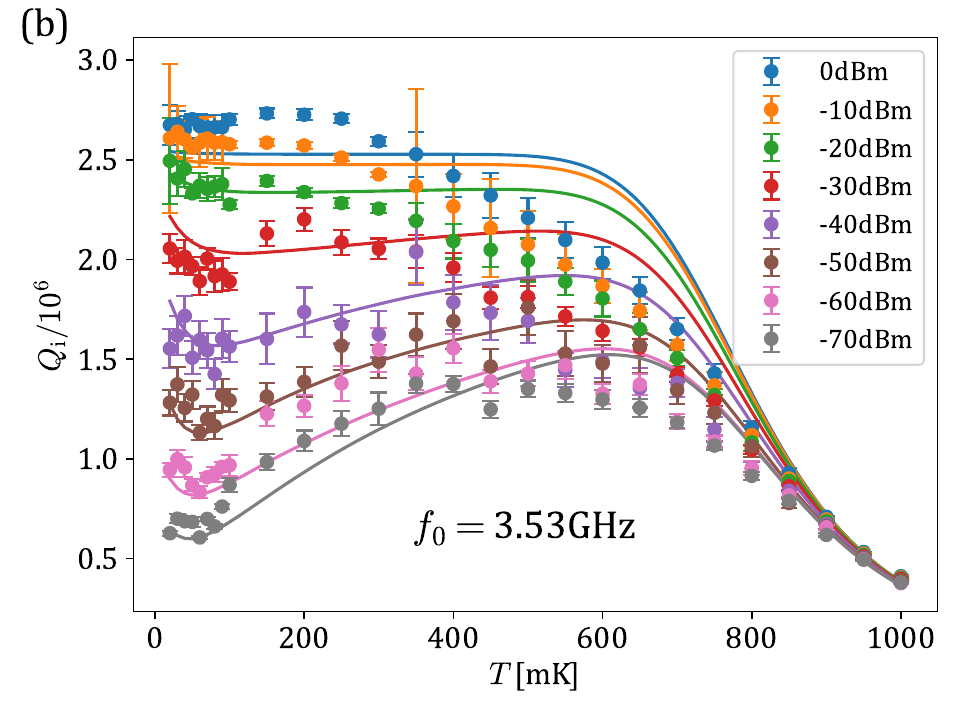}
        }\\
    \subfloat{
        \includegraphics[width=.48\linewidth, trim=0 15 0 0, clip]{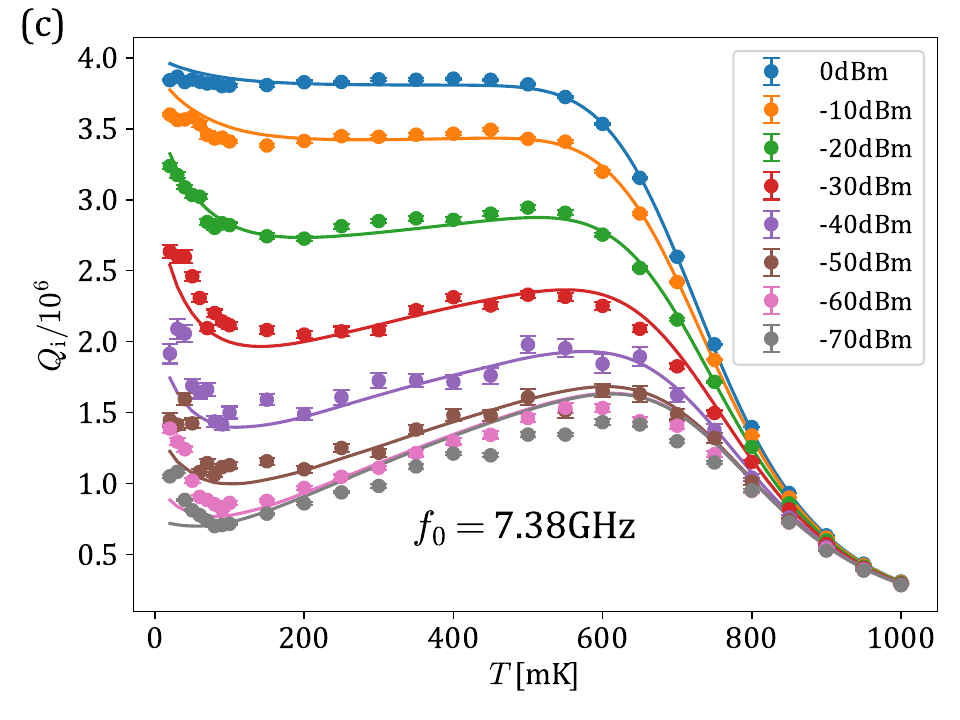}
        }\hfill
    \subfloat{
        \includegraphics[width=.48\linewidth, trim=0 15 0 0, clip]{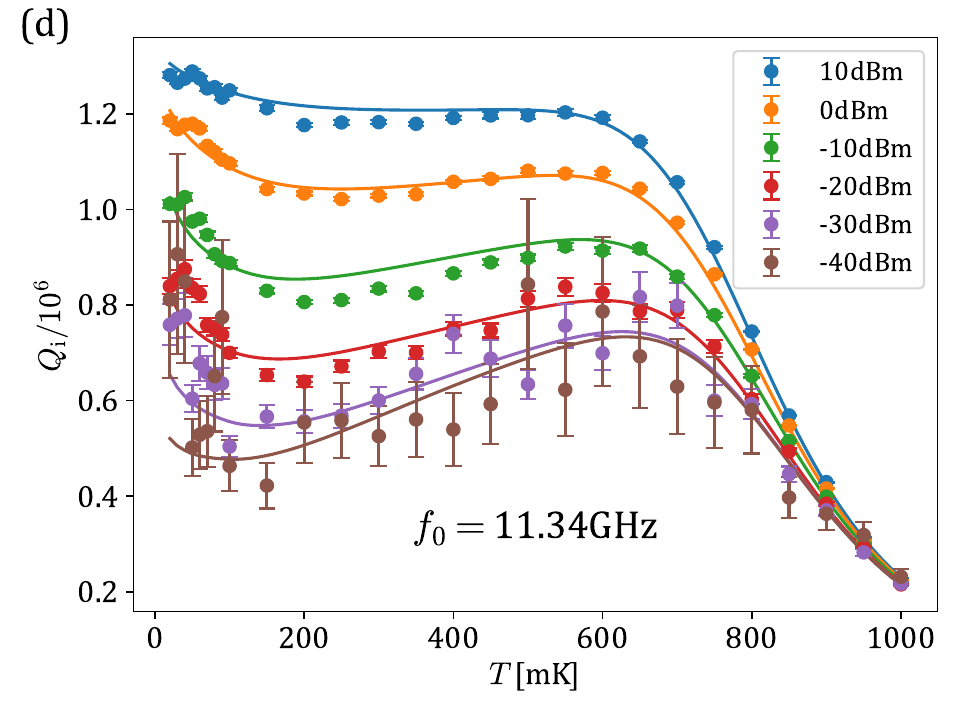}
        }\\
    \subfloat{
        \includegraphics[width=.48\linewidth, trim=0 10 0 0, clip]{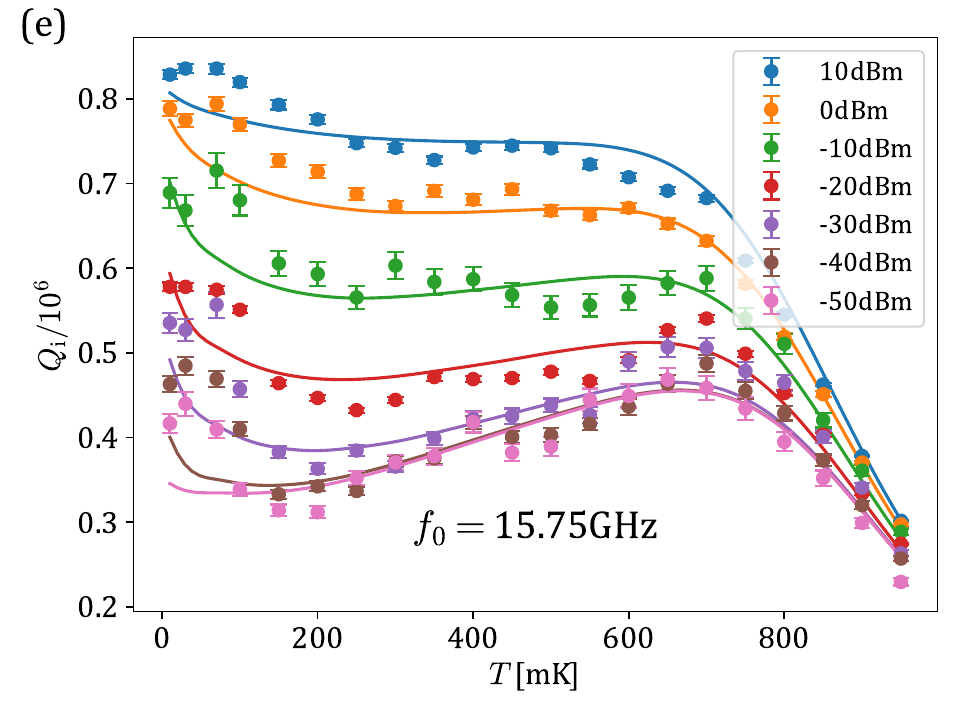}
        }\hfill
    \caption{Measurements of $Q_i$ in all resonators as a function of temperature and readout power, with simultaneous fits for each resonator. For clarity, the power is given at the entrance to the cryostat, although the fitting was done with the nominal power inside the resonators. The nominal frequencies of each resonator are written in the insets.}
    \label{fig:Qi_fit}
\end{figure*}

We present $Q_i$ measurements for all resonators at various temperatures and readout powers in Fig.~\ref{fig:Qi_fit}, with a fit of each resonator to our model. We see that the model captures the behavior quite well over many orders of magnitude of readout power and up to the critical temperature of tantalum. Measurements above $1K$ were not taken, but this regime is dominated by QPs, which are well described by the model. As the temperature decreases, the TLS loss becomes larger than the QP loss, which causes the gradual saturation of $Q_i$. In aluminum, the turnover temperature between TLS-dominated and QP-dominated loss is near 250mK~\cite{tai2024anomalous}, while our results show that for tantalum it is around 700mK. This is comparable to the turnover temperature measured in~\cite{crowley2023disentangling} which was about 600mK. At very low temperatures, a sharp increase in $Q_i$ is observed, which is due to an increase in the TLS saturation field $E_{\mathrm{s}}$, as seen in Eq.~\ref{eq:tan_deltaj}. This effect is less pronounced at high readout powers, as expected. The ability of the model to capture this phenomenon with a small number of free parameters is indicative of its physical validity. However, at extremely low readout powers, the model does not accurately capture the sharpness of this spike, meaning that the increase in saturation is larger than expected. Deviations also occur at higher temperatures for the low-power measurements. A possible explanation put forth in~\cite{tai2024anomalous} is the discreteness of the TLSs, as opposed to the continuous spectrum assumed in our model. Supporting this explanation are our results of the 3.53GHz resonator, as explained in the next paragraph.

For the 3.53GHz resonator, the turnover into the TLS saturation regime shown in Fig.~\ref{fig:Qi_fit}b is more gradual than can be explained by the model. We propose that this is due to the presence of a single TLS which is strongly coupled to the resonator mode. The contribution of a single coherent TLS with a low jump rate into the total Q factor is~\cite{tai2024anomalous}
\begin{equation} \label{eq:Q_single_TLS}
    Q_{\mathrm{TLS}}^{(1)} = \frac{Q_0^{(1)}}{\omega_{\mathrm{res}}} \frac{\Gamma_2^2(1+\kappa)+(U_{\mathrm{TLS}}/\hbar-\omega_{\mathrm{res}})^2}{\Gamma_2 \tanh{(U_{\mathrm{TLS}}/2k_{\mathrm{B}} T)}},
\end{equation}
where
\begin{equation} \label{eq:Q_single_TLS_defs}
\begin{split}
    &\Gamma_2 = \frac{\Gamma_1}{2} + 0.001 \frac{k_{\mathrm{B}} T}{\hbar}, \\
    &\Gamma_1 = \frac{C_0}{\hbar} \frac{U_{\mathrm{TLS}}}{\tanh{(U_{\mathrm{TLS}}/2k_{\mathrm{B}} T)}}, \\
    &\kappa = D_0 \frac{V_0^2}{\Gamma_1 \Gamma_2},
\end{split}    
\end{equation}
with $Q_0^{(1)}$, $U_{\mathrm{TLS}}$, $C_0$, and $D_0$ being model parameters. Notice that $D_0$ is not dimensionless. $U_{\mathrm{TLS}}$ is the TLS transition energy, $\Gamma_{1,2}$ are the TLS relaxation rates, and $\omega_{\mathrm{res}}$ is the angular resonant frequency of the resonator. 
We now give an approximate analysis of Eq.~\ref{eq:Q_single_TLS} to show how it can create the gradual turnover region observed in Fig.~\ref{fig:Qi_fit}b. 

At temperatures much lower than the TLS transition energy, $k_{\mathrm{B}} T\ll U_{\mathrm{TLS}}$, we see from Eq.~\ref{eq:Q_single_TLS_defs} that $\Gamma_1$ is constant at $C_0 U_{\mathrm{TLS}}/\hbar$. As the temperature increases $\Gamma_1$ also increases due to the onset of thermal saturation. However, in the regime $C_0\ll 0.002k_{\mathrm{B}} T/U_{\mathrm{TLS}}$ the second term in the expression for $\Gamma_2$ dominates, therefore $\Gamma_2\approx 0.001k_{\mathrm{B}} T/\hbar$. Eq.~\ref{eq:Q_single_TLS} becomes
\begin{equation} \label{eq:Q_single_TLS_approx}
\begin{split}
Q_{\mathrm{TLS}}^{(1)} \approx & \frac{Q_0^{(1)}}{\omega_{\mathrm{res}}} \frac{1}{\tanh{(U_{\mathrm{TLS}}/2k_{\mathrm{B}} T)}} \cdot \\
& \;\;\;\;\;\;\;\;\;\;\; \left( 0.001\frac{k_{\mathrm{B}} T}{\hbar} + \frac{(U_{\mathrm{TLS}}/\hbar-\omega_{\mathrm{res}})^2}{ 0.001\frac{k_{\mathrm{B}} T}{\hbar}} \right),    
\end{split}
\end{equation}
where we neglected the power dependence ($\kappa\approx 0$) for simplicity. We see from the second term in Eq.~\ref{eq:Q_single_TLS_approx} that at low temperatures it is possible to get a gradual decrease in $Q_{\mathrm{TLS}}^{(1)}$ with temperature before getting an increase at higher temperatures. For this to explain the gradual decrease measured in the 3.53GHz resonator, the turnover temperature $T^*$ between the region with decreasing $Q_{\mathrm{TLS}}^{(1)}$ and the region with increasing $Q_{\mathrm{TLS}}^{(1)}$ must be sufficiently high. This is only possible for $U_{\mathrm{TLS}} \gg k_{\mathrm{B}} T^*$, since otherwise the $\tanh(U_{\mathrm{TLS}}/2k_{\mathrm{B}} T)$ term in the denominator of Eq.~\ref{eq:Q_single_TLS_approx} will reduce the turnover temperature. Under this assumption, the  minimum point is at $k_{\mathrm{B}} T^*=1000(U_{\mathrm{TLS}}-\hbar\omega_{\mathrm{res}})$. Our measurements show $T^*\gg600\mathrm{mK}$ so we need $U_{\mathrm{TLS}}/h\gg3.5\mathrm{GHz}$. This means that the single TLS must be significantly detuned from the resonator for the gradual decrease to occur, justifying our assumption for $T\ll U_{\mathrm{TLS}}/2k_{\mathrm{B}}$. However, this detuning reduces the coupling of the TLS to the resonator. A possible solution is that the single TLS has a sufficiently large participation ratio with the resonator, so its effect is still significant. For large detunings, the Q-factor contribution of a single TLS coupled to a resonator is given by
\begin{equation}
    Q=\frac{\omega_{\mathrm{res}}}{\Gamma_2}\frac{\Delta^2}{g^2},
\end{equation}
where $\Delta$ is the TLS detuning from the resonator in Hz. We can assume $Q\approx 2\cdot 10^6$ and $\Gamma_2/2\pi\approx 10^{-3}k_{\mathrm{B}} T^*/h \approx 2 \mathrm{MHz}$, but to get an estimate for $\Delta$ a fit must be performed to find $U_{\mathrm{TLS}}$. 

\begin{figure}[htp!]
    \centering
    \includegraphics[width=\linewidth, trim=0 10 0 0, clip]{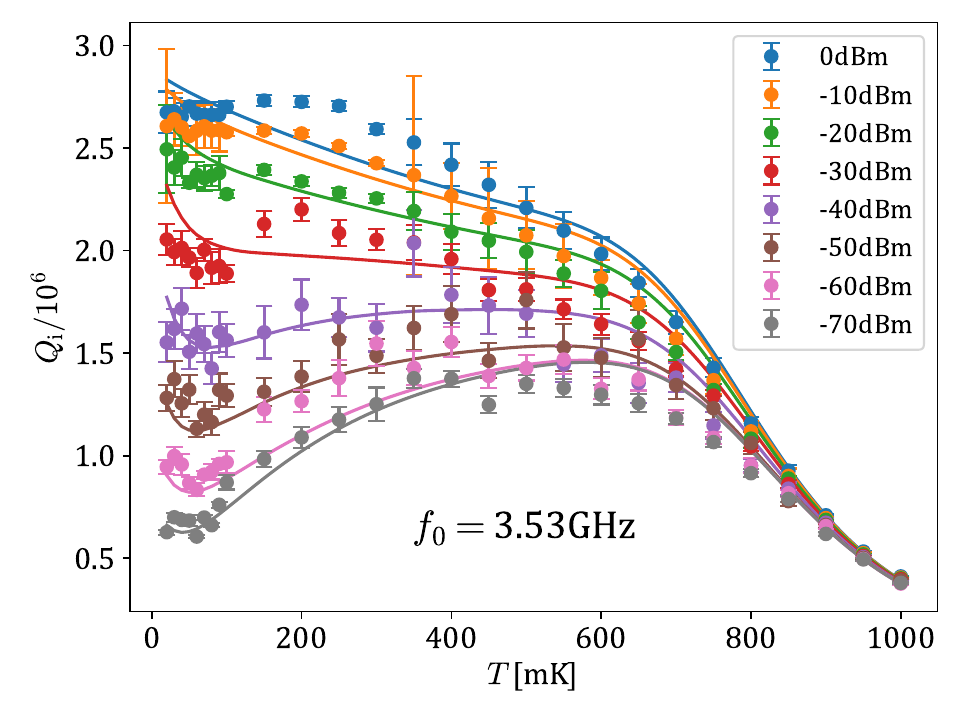}
    \caption{Fitting the 3.53GHz resonator with a model including the contribution from a strongly coupled single TLS at 40GHz. The gradual decrease in Q with temperature around 600mK is captured. The fit result gave $C_0=1.7\cdot 10^{-5}$, which satisfies the condition  $C_0\ll 0.002k_{\mathrm{B}} T/U_{\mathrm{TLS}}$ for the validity of Eq.~\ref{eq:Q_single_TLS_approx}, with $T=600\mathrm{mK}$ and $U_{\mathrm{TLS}}/h=40\mathrm{GHz}$}
    \label{fig:Qi_fit_n1_TLS}
\end{figure}

We performed a fit for the data of the 3.53GHz resonator, using a model that includes a strongly coupled single TLS at 40GHz according to Eq.~\ref{eq:Q_single_TLS} (i.e., without the above approximations). 40GHz was chosen because it was the minimal TLS frequency that could reproduce the data faithfully. This gives an overall 7-fit parameter model. The results are shown in Fig.~\ref{fig:Qi_fit_n1_TLS}. The fit is dramatically improved, supporting the plausibility of the strongly coupled far-detuned single TLS explanation. For this value of $U_{\mathrm{TLS}}$ we obtain $g\approx 1.09\mathrm{GHz}$, which is extremely high but not impossible. This very strong participation can explain why this effect is so rare, as despite the existence of many TLSs with a far detuning, almost no TLS has such a strong coupling to the resonators.

\begin{figure}[htp!]
    \centering
    \includegraphics[width=\linewidth, trim=0 0 0 0, clip]{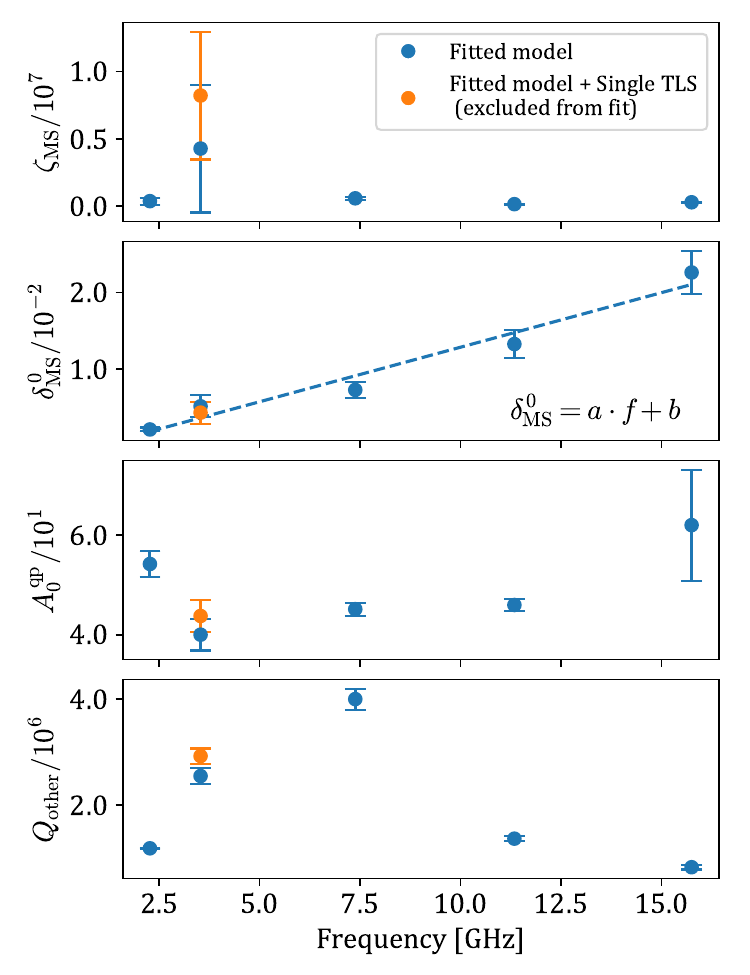}
    \caption{Best-fit parameters for the various resonators as a function of frequency. A clear trend emerges for $\delta_{\mathrm{MS}}^0$, with the empirical fitting functions shown in the insets. The units of $\zeta_{\mathrm{MS}}$ are $\mathrm{K^2/(V/m)^2}$ (SI units), while the other parameters are dimensionless. For $\zeta_{\mathrm{MS}}$, the 3.53GHz resonator data deviates significantly from the other resonators, and has larger uncertainty, due to its behavior being  not well described by our model. As detailed in the text, this is likely due to a strongly coupled single TLS.}
    \label{fig:params_vs_freq}
\end{figure}

Our aim in designing a wide span of frequencies for the resonators was to find empirical trends of the best-fit parameters with frequency. This is shown in Fig.~\ref{fig:params_vs_freq}. The uncertainties in the parameters were estimated by finding the range around the best values that produces an increase of 5\% in the square error of the fit. The standard $\chi ^2$ method was not applicable because the fitting function does not describe the data sufficiently well, and therefore the deviations from the fit do not distribute normally.


A clear linear trend with frequency in $\delta_{\mathrm{MS}}^0$ emerges, with parameters $a=0.145\cdot 10^{-2}$ns and $b=-0.151$. In the standard tunneling model of TLSs (STM) it is assumed that TLS-TLS interactions are much smaller than the typical random field exerted on each TLS, and as a result the single TLS density of states is roughly energy independent and proportional to $\delta_{\mathrm{MS}}^0$. This clearly does not match our results. However, other experiments also show a strong frequency dependence of $f^{\mu}$, where $0.1<\mu<0.3$~\cite{hunklinger1986thermal,stephens1973low,lasjaunias1975density} for the density of states, and even approaching $\mu\approx1$~\cite{skacel2015probing}. Such a significant dependence of the single TLS density of states on energy was shown to be a result of TLS-TLS interactions being of the order of the typical random field~\cite{churkin2021anomalous}, which may be a consequence of dominant electric dipole TLS-TLS interactions within the Two-TLS model~\cite{schechter2013inversion}. Our results support these findings and suggest that mechanisms beyond the STM play a crucial role in TLS loss in superconducting resonators. All other parameters of the model (namely $\zeta_{\mathrm{MS}}$, $Q_0^{\mathrm{qp}}$ and $Q_{\mathrm{other}}$) show no clear frequency dependence. This is expected because these parameters originate from fabrication non-uniformity in the chip and coupling to the electromagnetic environment, which are not related to the resonator frequency. \\

In this work, we have presented a simple model that enables systematic investigation of the contributions of different loss mechanisms in superconducting resonators. The core of the model is the TLS loss part, which is based on the STM~\cite{phillips1987two} but takes into account the spatial distribution of the electric field in the resonator structure. The motivation for this was that the level of TLS saturation depends on the local electric field intensity, which varies across the resonator. To obtain a complete loss model for the superconducting resonators, we combined the TLS loss model with QP loss and other loss mechanisms that do not depend on temperature or readout power, such as radiation loss. We performed experiments with tantalum CPW $\lambda/4$ transmission line resonators on a sapphire substrate, a platform that shows great potential for the fabrication of high-coherence superconducting circuits~\cite{place2021new,wang2022towards}. Our model was able to accurately describe the measurements across a wide range of readout powers and temperatures, and we extracted the model parameters for each resonator using a least-squares fit. Thanks to the wide frequency range of the measured resonators, we were able to examine frequency trends of the model parameters. We found a linear increase with frequency of $\delta^0_{\mathrm{MS}}$, which is proportional to the density of states of the TLS bath. This does not match the STM, which asserts a frequency-independent density of states, but does add to a growing body of experimental evidence indicating an increase in the density of states with frequency~\cite{hunklinger1986thermal,stephens1973low,lasjaunias1975density,skacel2015probing}. For the 3.53GHz resonator, the temperature dependence of the quality factor was more gradual than can be explained by the model. We attribute this to the presence of a single TLS, which is far detuned from the resonator but has a very high participation ratio with it. By incorporating this single TLS into our model, we were able to greatly improve the fit to this resonator, supporting the validity of this explanation. This demonstrates the importance of considering the discrete nature of TLSs, as opposed to a continuous spectrum~\cite{tai2024anomalous}.

\bibliographystyle{apsrev4-2}
\bibliography{bib}

\setcounter{equation}{0}
\setcounter{figure}{0}
\setcounter{table}{0}
\setcounter{page}{1}
\makeatletter
\renewcommand{\theequation}{S\arabic{equation}}
\renewcommand{\thefigure}{S\arabic{figure}}
\renewcommand{\bibnumfmt}[1]{[S#1]}
\renewcommand{\citenumfont}[1]{S#1}

\onecolumngrid

\end{document}